\documentclass{elsart}

\usepackage{graphicx}
\usepackage{amssymb}

\begin{document}

\begin{frontmatter}

\title{Geometry Effects at Atomic-Size Aluminium Contacts}

\author{U.~Schwingenschl\"ogl\corauthref{cor1}},
\ead{Udo.Schwingenschloegl@physik.uni-augsburg.de}
\author{C.~Schuster}
\corauth[cor1]{Corresponding author. Fax: 49-821-598-3262}
\address{Institut f\"ur Physik, Universit\"at Augsburg, 86135 Augsburg,
Germany}

\begin{abstract}
We present electronic structure calculations for aluminium nanocontacts.
Addressing the neck of the contact, we compare characteristic geometries
to investigate the effects of the local aluminium coordination on the
electronic states. We find that the Al $3p_z$ states are very sensitive
against modifications of the orbital overlap, which has serious
consequences for the transport properties. Stretching of the contact
shifts states towards the Fermi energy, leaving the system instable
against ferromagnetic ordering. By spacial restriction, hybridization
is locally suppressed at nanocontacts and the charge neutrality is
violated. We discuss the influence of mechanical stress by means of
quantitative results for the charge transfer.
\end{abstract}

\end{frontmatter}

Atomic-size contacts can be prepared by means of scanning tunneling
microscopy \cite{pascual93} or break junction techniques
\cite{scheer97}. In each case, piezoelectric elements are used
to stretch a wire with a precision of a few picometers until finally
a single atom configuration is reached. Such contacts have attracted
great attention over the last couple of years, in particular
concerning their electrical transport properties. Since transport is
restricted to a small number of atomic orbitals at the contact,
conductance across nanocontacts strongly depends on the local
electronic structure. An atomic-size constriction accomodates only a
small number of conducting channels, which is determined by the
number of valence orbitals of the contact atom. The transmission of
each channel likewise is fixed by the local atomic environment.
For a review on the quantum properties of atomic-size conductors see
Agra\"it {\it et al.} \cite{agrait03}.

From the theoretical point of view, the electronic structure and
conductance of nanocontacts and nanowires has been studied by ab
initio band structure calculations. For aluminium contacts,
investigations of the electronic states have been reported in
\cite{zabala02,ribeiro03,delin04a,delin04b}, and the conductance has
been addressed in
\cite{levy97,cuevas98a,cuevas98b,kobayashi00,palacios02,thygesen03,lee03,okano04,sasaki04}.
In these studies
various geometries have been used, which are assumed to model the
local atomic structure of the contact in an adequate way. The
breakage of an aluminium contact has been simulated by means of
molecular dynamics calculations in \cite{hasmy01,hasmy05,pauly06},
i.e.\ on the basis of realistic structural arrangements. However,
despite such a large number of investigations, the literature lacks
satisfactory reflections about the interrelations between the
details of the crystal structure and the local electronic states at
the nanocontact. In the present letter we will deal with this point
by comparing characteristic contact geometries, including stretched
configurations.

In a previous work \cite{cpl432} we have demonstrated that
hybridization between Al $3s$ und $3p$ states is strongly suppressed
at aluminium nanocontacts due to directed bonds at the neck of the
contact. We therefore expect the system to be very sensitive against
modifications of the orbital overlap coming along with the specific
contact geometry. As a consequence, structural details are important
for the electrical transport, since hybridization effects can
play a critical role for transport properties of atomic-size contacts
and interfaces \cite{schmitt05,cpl,epl}. In particular, stretching
of the nanocontact alterates the chemical bonding and thus may lead
to unexpected electronic features. We will show that it is mandatory
to account for the very details of the contact geometry in order
to obtain adequate results from electronic structure calculations.

The band structure results presented subsequently are obtained within
density functional theory and the generalized gradient approximation.
We use the WIEN2k program package, a state-of-the-art
full-potential code based on a mixed lapw and apw+lo basis \cite{wien2k}.
In our calculations the charge density is represented by
$\approx$150000 plane waves and the exchange-correlation potential is
parametrized according to the Perdew-Burke-Ernzernhof scheme. Moreover,
the mesh for the Brillouin zone integration comprises between 75 and 102
points in the irreducible wedge. While Al $1s$, $2s$, and $2p$ orbitals
are treated as core states, the valence states comprise Al $3s$ and $3p$
orbitals. The radius of the aluminium muffin-tin spheres amounts to 2.6
Bohr radii.

Our calculations rely on two characteristic contact geometries, which
we introduce in the following. On the one hand, we address a
configuration where a single Al atom is connected to planar Al
units on both sides, each consisting of seven atoms in a hexagonal
arrangement with fcc [111] orientation. The central sites of these
planar units lie on top of the contact atom, thus giving rise to
linear $\sigma$-type Al-Al bonds along the $z$-axis. For this reason,
we call the first geometry under consideration the {\it linear
contact configuration}. The finite Al units are connected to Al
$ab$-planes of infinite extension, which enables us to apply periodic
boundary conditions. We note that the contact Al site in this linear
geometry, due to its two nearest neighbours, resembles the essential
structural features of atoms in a monostrand nanowire \cite{cpl432}.

On the other hand, we study an Al atom sandwiched between two
pyramidal Al electrodes in fcc [001] orientation, which we call the
{\it pyramidal contact configuration}. To be specific, the
contact Al site has four crystallographically equivalent nearest
neighbours on both sides, which prohibits $\sigma$-type Al-Al bonding
via the $3p_z$ orbitals along the $z$-axis. Whereas the second
pyramidal layer comprises nine atoms, the third layer off the contact
extends infinitely on account of periodic boundary conditions.

For both contact configurations, a convenient choice for the bond
lengths and bond angles is given by the bulk (fcc) aluminium values,
therefore by interatomic distances of 2.86\,\AA. Mechanical stress
can increase this bond length at the nanocontact, which we simulate
by interatomic distances of 3.95\,\AA\ for the linear and 3.62\,\AA\
for the pyramidal contact configuration. In both cases, only the bond
lengths between the contact Al site and its nearest neighbours are
changed with respect to the fcc setup. Structural relaxation of the
electrodes due to the elongated contact bonds plays a minor role.

\begin{figure}
\centering
\includegraphics[width=0.5\textwidth,clip]{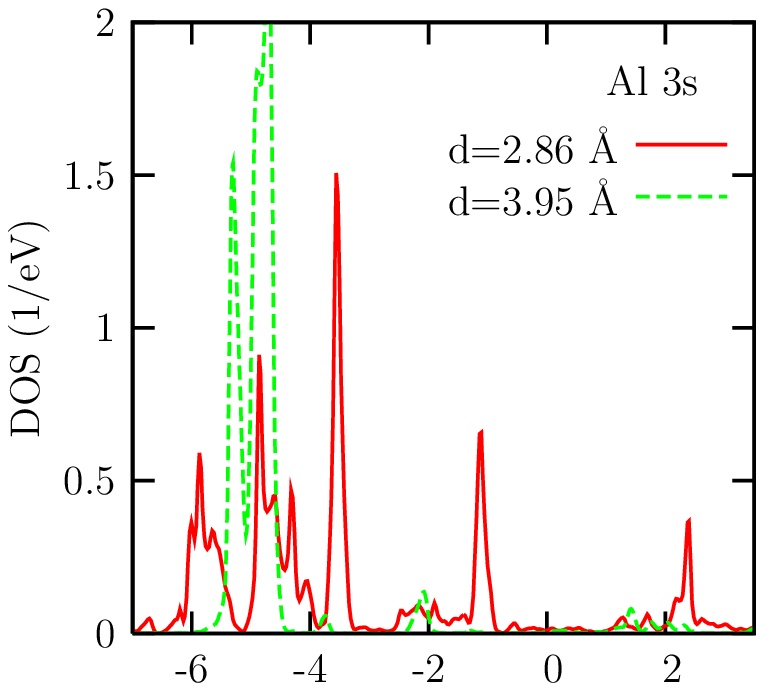}\\
\includegraphics[width=0.5\textwidth,clip]{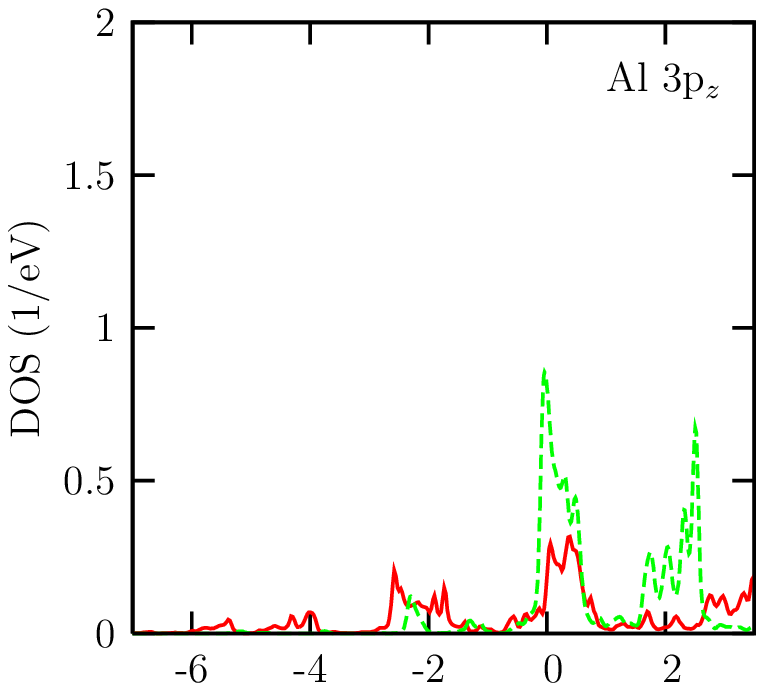}\\
\includegraphics[width=0.5\textwidth,clip]{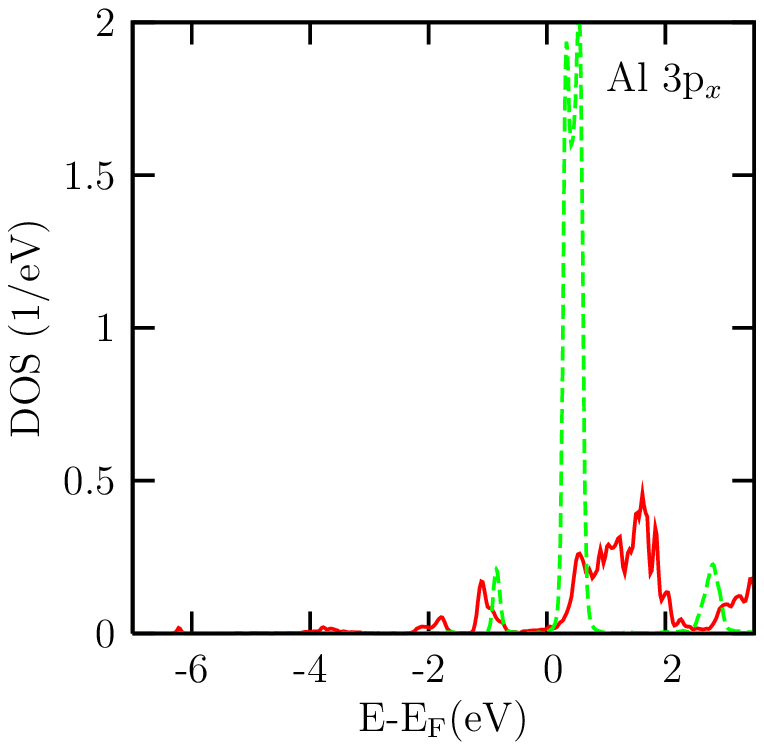}
\caption{Partial Al $3s$, $3p_z$, and $3p_x$ densities of states
for the contact aluminium site in the linear contact configuration.}
\label{fig1}
\end{figure}

For bulk aluminium it is well established that the formal Al
3$s^2$3$p^1$ electronic configuration is seriously interfered by
hybridization effects, giving rise to a prototypical $sp$-hybrid
system. However, the situation changes dramatically when covalent
bonding is no longer isotropic but restricted to specific directions,
as for an atomic-size contact. Partial Al $3s$, $3p_z$, and $3p_x$
densities of states (DOS) as calculated for the Al site at the neck
of our linear contact configuration are shown in figure\ \ref{fig1}.
By symmetry, $p_x$ and $p_y$ states are degenerate. While most of the
occupied states are of $3s$ type, the $3p$ states dominate at
energies above the Fermi level. Because hardly any contribution of
$3s$ and $3p$ states is found at energies dominated by the other
states, respectively, evolution of hybrid orbitals and an
interpretation in terms of $sp$-hybrid states is precluded. Most $3p$
electrons occupy the $p_z$ orbital, which is oriented along the
principal axis of the contact and therefore mediates $\sigma$-type
orbital overlap. Because neither $3s$ nor $3p_x$ states give rise to
significant contributions to the DOS at the Fermi energy, chemical
bonding is well characterized in terms of directed $3p_z$ bonds.

When the Al-Al bond length is stretched from 2.86\,\AA\ to 3.95\,\AA\
at the nanocontact, the central Al site decouples from its
neighbours. Its electronic states hence become more atom-like, which
is clearly visible for the $3s$ states in figure \ref{fig1}. Smaller
band widths and sharper DOS structures likewise are obvious for the
$3p_x$ states. Finally, for the $3p_z$ symmetry component we observe
a shift of states from lower energies to the Fermi level and from
higher energies to a new structure at about 2\,eV. The DOS at the
Fermi energy increases significantly.

Turning to the occupation of the valence orbitals, the decoupling of
the contact Al site comes along with a reduction of charge, amounting
to 0.13 electrons. In general, the $p_x$/$p_y$ atomic orbital does
not mediate chemical bonding due to the spacial restriction of the
crystal structure. Its occupation thus is strongly reduced and we
cannot expect local charge neutrality. Calculated values for the net
charge transfer off the contact Al site, as compared to bulk
aluminium, are given in table \ref{tab1} for bond lengths between
2.86\,\AA\ and 3.95\,\AA.

\begin{table}
\begin{tabular}{l|c|c}
configuration&bond length&charge transfer\\\hline
&&\\
linear&2.86\,\AA&0.51\\
&2.97\,\AA&0.54\\
&3.08\,\AA&0.57\\
&3.33\,\AA&0.59\\
&3.58\,\AA&0.62\\
&3.95\,\AA&0.63\\
&&\\
pyramidal&2.86\,\AA&0.22\\
&3.22\,\AA&0.53\\
&3.62\,\AA&0.61\\
&4.04\,\AA&0.66\\
\end{tabular}
\caption{\rm Net charge transfer off the contact aluminium site.}
\label{tab1}
\end{table}

\begin{figure}
\centering
\includegraphics[width=0.5\textwidth,clip]{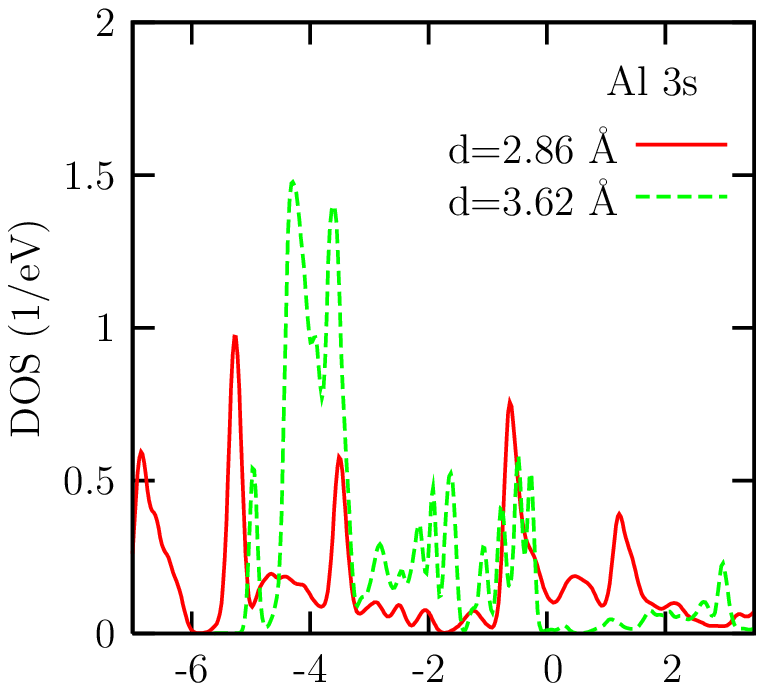}\\
\includegraphics[width=0.5\textwidth,clip]{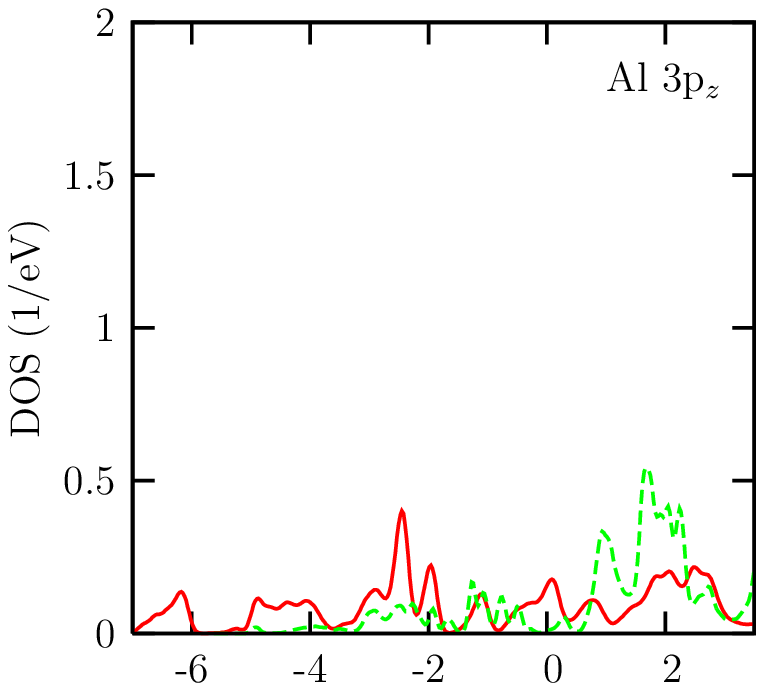}\\
\includegraphics[width=0.5\textwidth,clip]{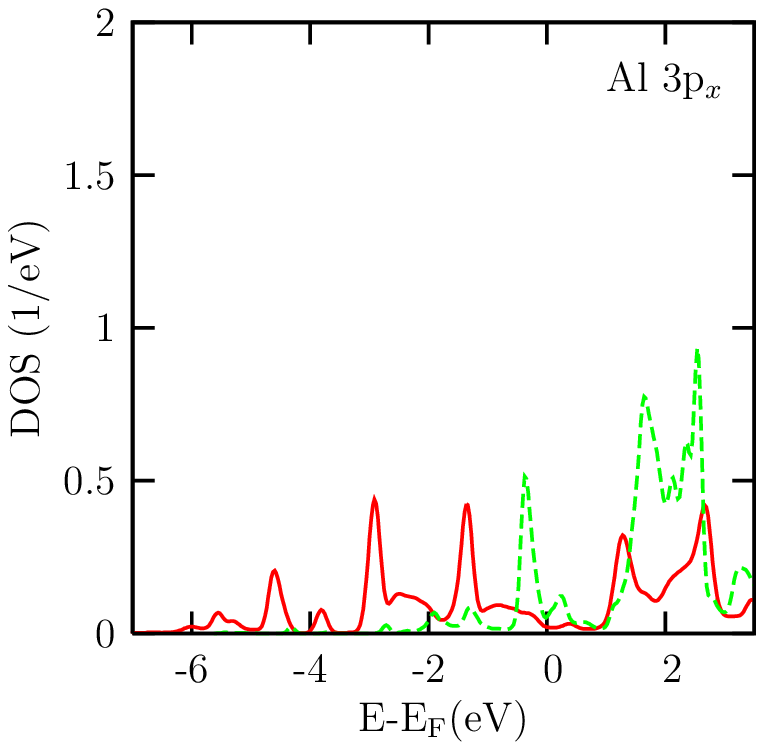}
\caption{Partial Al $3s$, $3p_z$, and $3p_x$ densities of states
for the contact aluminium site in the pyramidal contact configuration.}
\label{fig2}
\end{figure}

\begin{figure}
\centering
\includegraphics[width=0.5\textwidth,clip]{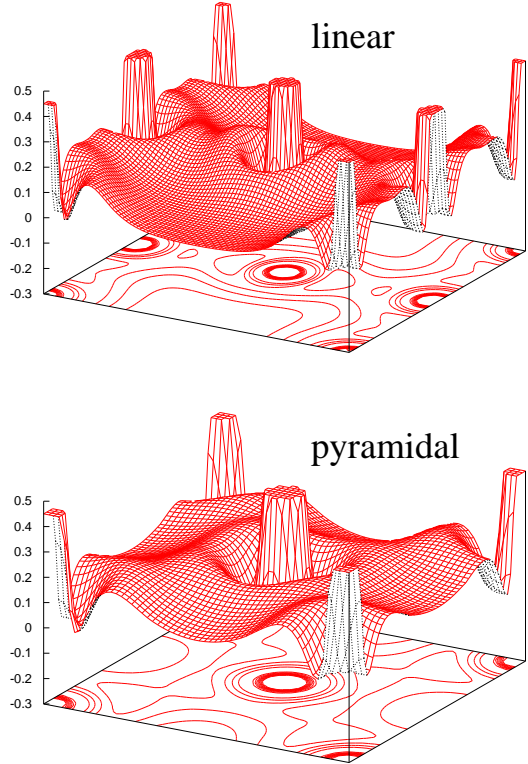}
\caption{Electron density maps for the neck of the linear and
pyramidal contact.}
\label{fig4}
\end{figure}

Figure \ref{fig2} shows partial Al $3s$, $3p_z$, and $3p_x$ densities
of states for the pyramidal contact configuration. As compared to the
linear case fundamental differences are found. For an Al-Al bond
length of 2.86\,\AA\ we now have a finite Al $3s$ DOS at the Fermi
energy. The same is true for the $3p_z$ states, whereas the $3p_x$
DOS almost vanishes. Again, $p_x$ and $p_y$ states are degenerate by
symmetry. While most of the occupied states still are of $3s$ type,
contributions of the $3p$ states are larger than for the linear
contact configuration. Chemical bonding therefore is more isotropic,
which is reflected by larger band widths. As expected from the
contact geometry, $\sigma$-type bonding via the $3p_z$-orbital is
significantly reduced. Due to increased hybridization between the
$3s$ and the three $3p$ orbitals, all these states can participate
in the electrical transport. Figure \ref{fig4} illustrates the
differences in the electronic structure at the linear and pyramidal
contact by means of electron density maps. In each case, the density
map covers the plane of the central Al site, where the principal axis
of the contact runs from left to right.

Stretching of the contact by increasing the Al-Al bond length
from 2.86\,\AA\ to 3.62\,\AA\ has similar effects as for the linear
contact configuration. The decoupling of the central Al site from
the pyramidal electrodes results in smaller band widths and sharper
DOS structures for the $3s$, $3p_z$, as well as $3p_x$ states, see
figure \ref{fig2}. In addition, the $3p_z$ und $3p_x$ states shift to
higher energies, giving rise to pronounced DOS structures near 2\,eV.
Due to the reduced band width, the $3s$ and $3p_z$ DOS disappears
almost completely in the vicinity of the Fermi energy. As concerns 
the $3p_z$ states, the pyramidal geometry thus shows the opposite
behaviour than the linear geometry. Since the Al $3p_z$ states
mediate the main part of the orbital overlap across the nanocontact,
they are very sensitive against changes in the crystal structure.
Structural rearrangement during the breakage of an aluminium 
nanocontact or nanowire consequently should have serious effects on
the electrical transport, such as modulation of the conductance.

Elongated bonds again are accompanied by a decline of charge at the
contact. However, the value of 0.39 electrons net charge transfer
off the central Al site is larger than in the linear case. This
traces back to the fact that reduction of hybridization on stretching
is more efficient for the pyramidal geometry, since for the linear
geometry hardly any hybridization is left right from the beginning.
Accordingly, smaller values for the net charge transfer are found in
the pyramidal case, see table \ref{tab1}.

\begin{figure}
\centering
\includegraphics[width=0.5\textwidth,clip]{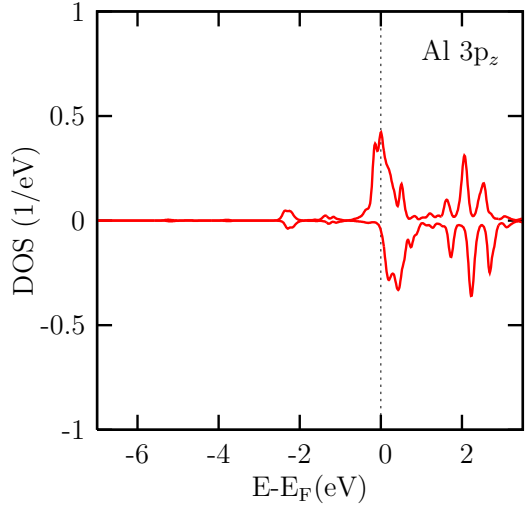}
\caption{Spin majority and minority Al $3p_z$ densities of states
for the contact aluminium site in the stretched linear contact
configuration ($d=3.95$\,\AA).}
\label{fig3}
\end{figure}

We next show that the remarkable increase of the Al $3p_z$ DOS at
the Fermi energy on stretching the linear contact configuration leads
to an instability against ferromagnetic ordering. A spontaneous
magnetization of simple metal nanowires has been predicted by Zabala
{\it et al.} \cite{zabala02}. These authors have demonstrated the
instability in explicit calculations for an aluminium nanowire, using
a stabilized jellium model. For Al-Al bond lengths of $d=3.95$\,\AA\
at the neck of our contact, we hence have performed spin polarized
electronic structure calculations, yielding a stable ferromagnetic
solution. Figure \ref{fig3} displays the corresponding spin majority
and minority densities of states for the contact Al site. The local
spin splitting of about 0.2\,eV is connected to an energy gain of
5.5\,mRyd. Moreover, the magnetic moment
amounts to 0.1\,$\mu_B$, which is significantly smaller than the
0.68\,$\mu_B$ reported by Zabala {\it et al.} \cite{zabala02} for an
infinite nanowire. However, Delin {\it et al.} \cite{delin04a} have
shown for Pd nanowires that the spontaneous magnetization
decreases rapidly for short chains. Only the linear contact
configuration is subject to a magnetic instability on stretching.
For the pyramidal geometry, of course, magnetism cannot be expected,
compare the spin degenerate DOS curves in figure \ref{fig2}.

In conclusion, we have studied the electronic structure of aluminium
nanocontacts by means of band structure calculations within density
functional theory. Taking into account the details of the crystal
structure, we have discussed the electronic features of prototypical
contact geometries (linear versus pyramidal). Our calculations result
in two largely different scenarios. In particular, the Al $3p_z$
states are strongly affected by modifications of the chemical
bonding. If $\sigma$-type bonding via the $3p_z$ orbitals is dominant
because of direct orbital overlap across the contact, $sp$-type
hybridization is almost completely suppressed and only Al $3p_z$
states remain at the Fermi energy \cite{cpl432}. Otherwise, if the bonding is more
isotropic for geometrical reasons, the Al $3s$ states likewise have
to be taken into account. The divers behaviour of the linear and
pyramidal geometry becomes even more pronounced when the contact is
stretched. Whereas in the linear case the Al $3p_z$ DOS at the Fermi
energy increases, which even yields a ferromagnetic instability, it
vanishes in the pyramidal case, leaving the system insulating. As a
consequence, the structural details of the contact are expected to
strongly influence the electrical transport. Because of
structural rearrangements, they are particularly relevant for the
breakage of a nanocontact or nanowire.

\section*{Acknowledgements}
We thank U.\ Eckern and P.\ Schwab for helpful discussions and the
Deutsche Forschungsgemeinschaft for financial support (SFB 484).

\end{document}